\documentclass[prx,twocolumn,amsmath,amssymb,superscriptaddress]{revtex4-1}

\usepackage{amsmath,amssymb,amsfonts,mathrsfs} 	
\usepackage{graphicx}
\usepackage{subfigure}
\usepackage{xcolor}
\usepackage[normalem]{ulem}
\usepackage[colorlinks=true,linkcolor=blue,citecolor=blue,urlcolor=black]{hyperref}

\newcommand{\ei}[1]{{\rm e}^{i #1}}
\newcommand{\emi}[1]{{\rm e}^{-i #1}}
\newcommand{\kk}{\mbox{\boldmath$\kappa$}}

\def\beq{\begin{equation}}
\def\eeq{\end{equation}}
\def\bea{\begin{eqnarray}}
\def\eea{\end{eqnarray}}
\def\nn{\nonumber\\}

\def\ket#1{\vert#1\rangle}

\def\me#1#2#3{\langle#1\vert#2\vert#3\rangle}
\def\ev#1{\langle#1\rangle}

\def\0{{\bf 0}}
\def\r{{\bf r}}
\def\v{{\bf v}}
\def\k{{\bf k}}

\def\p{{\bf p}}

\def\E{{\bf E}}

\def\EE{{\cal E}}

\def\EEE{\mbox{\boldmath${\cal E}$}}

\def\B{{\bf B}}
\newcommand{\intk}{\int_{\rm BZ} \! \frac{d {\bf k}}{(2\pi)^d} \;}
\renewcommand{\[}{\begin{equation}}
\renewcommand{\]}{\end{equation}}

\newcommand{\equ}[1]{Eq.~(\ref{#1})}
\newcommand{\eqs}[2]{Eqs.~(\ref{#1}) and (\ref{#2})}
 
\def\runtime{(\the\time)\qquad\the\month/\the\day/\the\year}
\def\today
 {\count10=\year\advance\count10 by -2000 \number\day--\ifcase
  \month \or Jan\or Feb\or Mar\or Apr\or May\or Jun\or
             Jul\or Aug\or Sep\or Oct\or Nov\or Dec\fi--\number\count10}

\def\hour{\count10=\time\count11=\count10
\divide\count10 by 60 \count12=\count10
\multiply\count12 by 60 \advance\count11 by -\count12\count12=0
\number\count10 :\ifnum\count11 < 10 \number\count12\fi\number\count11}

\begin{document}

\title{Theory of nonlinear dc conductivity, longitudinal and transverse}

\author{Raffaele Resta}

\affiliation{Istituto Officina dei Materiali IOM-CNR, Strada Costiera 11, 314151 Trieste, Italy}\email{Permanent addres; e-mail resta[at]iom.cnr.it}
\affiliation{Institute of Physics of the Czech Academy of Sciences, 18221 Prague 8, Czechia}
\affiliation{Donostia International Physics Center, 20018 San Sebasti{\'a}n, Spain}
\date{\today}

\begin{abstract} Kohn's theory of Drude conductivity, established in a many-body framework, addresses even systems with disorder and correlation, besides the ordinary band metals (i.e. crystalline systems of independent electrons). Kohn's theory is here extended to nonlinear dc conductivities of arbitrary order, longitudinal and transverse. The results are then reformulated in a band-structure framework, and their relationships to the semiclassical theory of nonlinear electron transport are elucidated.
\end{abstract}

\maketitle

\section{Introduction}

The publication in 2015 of the Sodemann-Fu paper \cite{Sodemann15} about the quadratic Hall conductivity drew much interest, both theoretical and experimental, on nonlinear conductivities in general, both Hall and longitudinal. Refs. \cite{Nandy19,Ma19,Du19,Kang19,Battilomo19,Zhou20,Ortix21,Watanabe20,Lahiri21,Tsirkin21,Zhang21,Du21,Zelezny21,Tanikawa21} are just some of the many papers and preprints devoted to nonlinear electron transport which appeared in the most recent years. With the exception of Ref. \cite{Watanabe20}, all of  the theoretical work in the quoted papers is set in a semiclassical framework.

Here I take a different path, by going back to Kohn's  time-honored theory of Drude conductivity \cite{Kohn64,Scalapino93,rap157}, and showing how it naturally extends to deal with nonlinear conductivities of any order, both Hall and longitudinal. Since dissipation cannot enter Schr\"odinger equation directly (at variance with Boltzmann equation), the response functions are causal but nondissipative, and the induced current does not reach a steady state. For instance it is well known that---to lowest order---a dc field induces in a pristine metal a longitudinal free acceleration of the many-electron system; extrinsic effects are needed to retrieve Ohm's law. At the simplest level, such effects are summarized into an heuristic relaxation time $\tau$. I am going to generalize this, by explicitly showing the time dependence of the higher-order induced currents, both Hall and longitudinal, in a pristine material. The selection rules dictated by time-reversal (T) and inversion (I) symmetries will also be discussed.

To linear order---and only to linear order---a dc field may induce a steady transverse current in a T-breaking pristine material; extrinsic effects are not needed, although they actually contribute to the effect in real materials \cite{Nagaosa10}. The intrinsic linear Hall conductivity is generally called ``geometrical'' \cite{Vanderbilt,rap149}; it becomes topological in a two-dimensional insulator \cite{Thouless82,Niu85,Xiao10}. In all of the other cases a steady current is reached only in presence of extrinsic disssipation mechanisms. 

Switching from the time domain to the $\omega$ domain makes straightforward to account for extrinsic effects at the simplest level: the nonlinear responses (to any order in the field) are converted from causal to dissipative by means of an heuristic $\tau$. The standard approach adopted in the well known case of Drude conductivity \cite{rap157} is generalized here to the longitudinal and Hall conductivities of any order. Their $\tau$-dependences is found to be qualitatively the same as for the semiclassical theories, based on Boltzmann equation \cite{Lahiri21,Tsirkin21,Zhang21}.

All of the expressions here obtained in a compact-many body formalism are then converted into their band-structure analogues, in order to address crystalline systems of noninteracting electrons. Only the metallic case is relevant, because all nonlinear conductivities vanish in insulators. A detailed comparison, based on the time evolution of the adiabatic current induced by a constant field, confirms that the semiclassical approach, in the $\tau \rightarrow \infty$ limit, provides indeed the same results---to all orders in the field---as the full quantum-mechanical approach at the band-structure level.

Sec. II displays the (by now famous) ``Hamiltonian with a flux'' as introduced by Kohn in 1964, and provides the related expression for the many-body current density, exploited in the following Sections. Sec. III collects all the results about dc linear conductivity: III-A presents the many body expression for the anomalous Hall conductivity, including its quantized version for a two-dimensional insulator; III-B presents an alternative derivation of Kohn's famous expression for the Drude weight; III-C introduces the concept of Born effective charges in metals, and generalizes to a many-body framework the sum rule obeyed by them, recently found by Dreyer, Coh, and Stengel within band-structure theory \cite{prep}. Sec. IV shows how the same logic as in Sec. III can be extended in order to deal with quadratic conductivity, both Hall and longitudinal. After the thorough discussion in Sec. IV, the following step of addressing an arbitrary order is quite straightforward: this is shown in Sec. V. The general many-body formalism is specialized in Sec VI to band metals and band insulators (i.e. crystalline systems of noninteracting electrons), where some of the known formul\ae\ are retrieved; the relationships between the nonlinear dc conductivities obtained from ab-initio band-structure theory and those obtained from  the semiclassical approach are discussed. Sec VI contains some concluding remarks. Finally, some technical developments have been expunged from the main text and confined into three Appendices.

\section{Electron transport in a many-body framework}

\subsection{Kohn Hamiltonian}

The starting point of the present theory is a milestone paper published by Kohn in 1964 \cite{Kohn64}. Following him, we consider a system of $N$ interacting $d$-dimensional electrons in a cubic box of volume $L^d$, and the family of many-body Hamiltonians parametrized by $\kk$, called ``flux'' or ``twist'': \[ \hat{H}_{\kk} = \frac{1}{2m} \sum_{i=1}^N \left(\p_i + \hbar \kk \right)^2 + \hat{V}, \label{kohn} \] where $\hat{V}$ includes the one-body potential (possibly disordered) and electron-electron interaction.
We assume the system to be macroscopically homogeneous; the eigenstates $\ket{\Psi_{n\kk}}$ are normalized to one in the hypercube of volume $L^{Nd}$. 
The thermodynamic limit $N \rightarrow \infty$, $L \rightarrow \infty$, $N/L^d=n$ constant is understood throughout this work. In order to simplify notations I will set $\hat{H}_{0} \equiv \hat{H}$, $\ket{\Psi_{n0}} \equiv \ket{\Psi_{n}}$ , $E_{n0} \equiv E_n$.

We assume Born-von-K\`arm\`an (BvK) periodic boundary conditions: the many-body wavefunctions are periodic with period $L$ over each electron coordinate $\r_i$ independently; the
potential $\hat{V}$ enjoys the same periodicity. The flux $\kk$---cast into inverse-length dimensions for convenience---corresponds to perturbing the Hamiltonian with a vector potential  $\hbar c \kk /e$, constant in space. While Kohn only considered a time-independent $\kk$, here I consider instead an adiabatically  time-dependent flux, which amounts to perturbing the Hamiltonian with the macroscopic field $\EEE(t) = - \hbar \dot\kk(t)/e$. The electron response will be evaluated by means of $\kk$ derivatives at $\kk=0$, and we notice that for $\kk=0$ the Hamiltonian of \equ{kohn} is T-invariant. Following Kohn, $\kk$-derivatives must be evaluated first, and the $L \rightarrow \infty$ limit taken afterwards \cite{Kohn64,Scalapino93}.

A T-breaking modification of Kohn's Hamiltonian is \[ \hat{H}_{\kk} = \frac{1}{2m} \sum_{i=1}^N \left[\p_i + \frac{e}{c}{\bf A}(\r_i) + \hbar \kk \right]^2 + \hat{V}, \label{kohn2} \] where the vector potential summarizes all intrinsic T-breaking terms, as e.g. those due to a coupling to a background of local moments; in this case ${\bf A}(\r)$ enjoys BvK periodicity. The vector potential could even account for a macroscopic $\B$ field, provided that it is commensurate, and that the BvK boundary conditions are modified accordingly \cite{Niu85}.

\subsection{Macroscopic current}

The kinetic-energy term in \equ{kohn2} defines the extensive many-electron velocity operator as \[ \hat{\v}_{\kk} = \frac{1}{m} \sum_{i=1}^N \left[\p_i + \frac{e}{c}{\bf A}(\r_i) + \hbar \kk \right] = \frac{1}{\hbar} \partial_{\kk} \hat{H}_{\kk} . 
\] When $\kk$ is adiabatically varied in time the instantaneous current density is the sum of two terms: the expectation value of the current operator, and the Niu-Thouless adiabatic current \cite{Niu84,Xiao10}. Their expression is cast as:
\bea  j_\alpha &=& - \frac{e}{ \hbar L^d} \me{\Psi_{0\kk}}{\partial_{\kappa_\alpha} \hat{H}_{\kk}}{\Psi_{0\kk}} \nn &+& \frac{ie}{L^d}( \ev{\partial_{\kappa_\alpha}{\Psi}_{0\kk} |  \dot{\Psi}_{0\kk} } - \ev{\dot{\Psi}_{0\kk} | \partial_{\kappa_\alpha} \Psi_{0\kk} } ) \nn  &=& - \frac{e}{\hbar L^d} \partial_{\kappa_\alpha} E_{0\kk} +  \frac{e}{L^d}\Omega_{\alpha\beta}(\kk) \dot{\kappa}_\beta \, ,  
\label{currentx} \eea where the sum over repeated Cartesian indices is understood, and $\Omega_{\alpha\beta}(\kk)$ is the many-body Berry curvature \[ \Omega_{\alpha\beta}(\kk)  = -2 \,\mbox{Im } \ev{\partial_{\kappa_\alpha} \Psi_{0\kk} | \partial_{\kappa_\beta} \Psi_{0\kk}} . \] We consider from now on only the adiabatic response to a field constant in time, in which case $\kk = \kk(t) = - et \EEE/\hbar$. The macroscopic current to all orders in $\kk$---ergo to all orders in $\EEE$ and in $t$---is: \[ j_\alpha(t) = - \frac{e}{\hbar L^d} \partial_{\kappa_\alpha} E_{0\kk} - \frac{e^2}{\hbar L^d}\Omega_{\alpha\beta}(\kk) \EE_\beta . 
\label{current} \]
The extensive quantity $-\Omega_{\alpha\beta}(\kk) \dot{\kappa}_\beta$ is the many-electron anomalous velocity, normal to the electric field $\EEE$: the second term in \equ{current} accounts therefore for a purely transverse current, which can be nonzero in either insulators or metals. 

The first term in \equ{current} is not new: it already appeared in this form in Ref. \cite{Watanabe20}, where it is exploited in a somewhat different way from the present one. This term accounts for a current which in general is not parallel to the field. To lowest order this term yields the symmetric part of the dc linear-conductivity tensor, ergo is by definition longitudinal; when dissipation is accounted for, the current is Ohmic. 
Beyond the linear regime and in a {\it macroscopic} approach the partition of the current into Hall and Ohmic components becomes subtle in low-symmetry situations \cite{Tsirkin21}. In the present {\it microscopic} theory the partition in two terms is unambiguous: I am going to call throughout as ``longitudinal'' the current from the first term in \equ{current}, labeled with a superscript ``(+)''; the second term will originate ``Hall' currents, labeled with ``(-)''

In insulators the expectation value of the many-body velocity is zero to any order in $\EEE$---bar a dielectric breakdown---ergo the longitudinal current vanishes and  $E_{0\kk}$ is $\kk$-independent; in metals instead $E_{0\kk}$ actually depends on $\kk$, because periodic boundary conditions violate gauge-invariance in the conventional sense \cite{Kohn64}.    

\section{Linear conductivity}

The conventional setting for linear conductivity is in the $\omega$ domain: \[ j_\alpha(\omega) = \sigma_{\alpha\beta}^{(+)}(\omega) \,\EE_\beta(\omega) + \sigma_{\alpha\beta}^{(-)}(\omega) \, \EE_\beta(\omega), \] where the symmetric (longitudinal) and antisymmetric (Hall) components of the conductivity tensor are explicitly separated.

At finite $\omega$ the tensor obtains from time-dependent linear-response theory, via the appropriate Kubo formul\ae; here instead we focus on the dc components only, and we address them  by means of the adiabatic response of the many-electron system. The appeal of the present approach is that it can be very naturally generalized to nonlinear conductivities. For the sake of completeness, Appendix \ref{sec:kubo} reports the conventional derivation of the linear results given below via Kubo formul\ae, for both the Hall and longitudinal cases. 

\subsection{Hall conductivity}

By setting $\kk=0$ in the second term of \equ{current} we get immediately the linear Hall tensor: \[  \sigma_{\alpha\beta}^{(-)}(0) = - \frac{e^2}{\hbar L^d}\Omega_{\alpha\beta}(0) ; \label{hall} \] 
the expression holds for either insulators or metals, for either $d=2$ or $d=3$, and yields the geometric (or intrinsic) term in the Hall conductivity \cite{Nagaosa10}; it can be nonzero only if the Hamiltonian breaks  T-symmetry at $\kk=0$ (see also the discussion below about symmetry).  

In the special case of an insulator and $d=2$ \equ{hall} is quantized in the large-$L$ limit:
 \[ \sigma_{xy}^{(-)}(0) = - \frac{e^2}{h} C_1 ,\label{ch2} \] where $C_1\in {\mathbb Z}$ is a Chern number. This famous relationship was first established at the independent-electron level, where $C_1$ is also known as TKNN invariant \cite{Thouless82}; it was later generalized by Niu, Thouless, and Wu, who provided the many-body expression for $C_1$ \cite{Niu85}. Following Ref. \cite{Xiao10} (Sec. III.C) the same invariant is conveniently recast as \[ C_1 = \frac{1}{2\pi} \int_0^{\frac{2\pi}{L}} \!\! d\kappa_x \int_0^{\frac{2\pi}{L}} \!\! d\kappa_x  \;  {\sf \Omega}_{xy}(\kk) \label{chern} ; \] \equ{chern} is quantized because it is equivalent to the integral over a torus. 
 
In order to retrieve this result within the present approach, we focus on the simple case of no macroscopic ${\bf B}$ field, i.e. on the so called quantum anomalous Hall effect. I start reminding
that in insulators the ground-state energy $E_{0\kk}$ is $\kk$-independent, and I define $\hat{\r} = \sum_I \r_i$. I then
observe that
whenever the components of $\kk - \kk'$ are integer multiples of $2\pi/L$, then the state $\ei{(\kk-\kk')\cdot \hat{\r}} \ket{\Psi_{0\kk}}$ 
obeys both the Schr\"odinger equation and BvK boundary conditions, ergo
is eigenstate of $\hat{H}_{\kk'}$ with the same eigenvalue as $\ket{\Psi_{0\kk}}$. The eigenstates which define ${\sf \Omega}_{xy}(\kk)$ have therefore the required toroidal periodicity: \[ \ket{\Psi_{0\kk'}} = \ei{(\kk-\kk')\cdot \hat{\r}} \ket{\Psi_{0\kk}} . \label{gauge} \] Since ${\sf \Omega}_{xy}(\kk)$ is gauge-invariant, an arbitrary $\kk$-dependent phase factor may relate the two members of \equ{gauge}. It is worth stressing that in the topological case a globally smooth periodic gauge does not exist and an ``obstruction'' is necessarily present; in other words one can enforce \equ{gauge} as it stands (with no extra phase factor) only locally, not globally; we also notice that \equ{gauge} may be regarded as the many-body analogue of the periodic gauge in band-structure theory \cite{Vanderbilt}.

\equ{chern} is independent of the $L$ value, and its integrand is extensive: therefore in the large-$L$ limit the integration domain contracts to a point: \[ C_1 = \frac{1}{2\pi} \left(\frac{2\pi}{L} \right)^2  \Omega_{xy}(0) .  \label{single} \] By comparing this to \equ{hall} for $d=2$, \equ{ch2} is immediately retrieved. 

Finally, it is worth stressing that \equ{single}---at variance with \equ{chern}---is {\it not} quantized at finite $L$: it only becomes  quantized in the $L \rightarrow \infty$ limit. Indeed the convergence with $L$ of the single-point Chern number, \equ{single}, has been investigated long ago by actual simulations based on an independent-particle model Hamiltonian: see Fig. 2 in Ref. \cite{rap135}.

\subsection{Drude conductivity} \label{sec:drude}

The longitudinal response linear in the field obtains by taking the time derivative  of the first term in \equ{current}: \bea \partial_t j^{(+)}_\alpha(t) &=& - \frac{e}{\hbar L^d} \frac{\partial^2 E_{0\kk}}{\partial t \,\partial \kappa_\beta}  \label{rapix}\\ &=& - \frac{e}{\hbar L^d} \frac{\partial^2 E_0}{\partial \kappa_\alpha \partial \kappa_\beta} \dot\kappa_\beta = \frac{e^2}{\hbar^2 L^d} \frac{\partial^2 E_0}{\partial \kappa_\alpha \partial \kappa_\beta} \EE_\beta . \nonumber  \eea This derivative is time-independent, ergo the many-electron system undergoes free acceleration; \equ{rapix} can be recast as \[ \partial_t j^{(+)}_\alpha = \frac{D_{\alpha\beta}}{\pi} \EE_\beta , \qquad D_{\alpha\beta} = \frac{\pi e^2}{\hbar^2 L^d} \frac{\partial^2 E_0}{\partial \kappa_\alpha \partial \kappa_\beta} \label{rapix2} , \] where $D_{\alpha\beta}$ is the Drude weight, as defined by Kohn \cite{Kohn64,Scalapino93,rap157}; it clearly measures the inverse inertia of the many-electron system when probed by a constant macroscopic field.

In the $\omega$ domain the Drude conductivity is expressed as
$ j^{(+)}_\alpha(\omega) = \sigma^{(+)}_{\alpha\beta}(\omega) \EE_\beta$, \[ \sigma^{(+)}_{\alpha\beta}(\omega) = \frac{D_{\alpha\beta}}{\pi} \frac{i}{\omega + i\eta} = D_{\alpha\beta} \left[ \delta(\omega) + \frac{i}{\pi \omega} \right] , \label{omega1} \] where the positive infinitesimal $\eta$ ensures causality \cite{rap157}; an alternative derivation is provided in Appendix \ref{sec:freq}. 
The $\delta(\omega)$ singularity is a fingerprint of the free acceleration in the time domain. Notice that only the dc contribution to $\sigma^{(+)}_{\alpha\beta}(\omega)$ is considered here and could be derived from the adiabatic response; the full tensor requires time-dependent perturbation theory and is dicussed in Appendix \ref{sec:kubo}.    

The response functions as considered so far are causal but nondissipative: one cannot indeed include dissipation within the Schr\"odinger equation of motion. Nonetheless dissipation can be inserted {\it a-posteriori} via a relaxation time $\tau$, by heuristically replacing $\partial_t$ with $\partial_t + 1/\tau$ in \equ{rapix2}, and then setting the $\partial_t$ term equal to zero (steady state); equivalently, one could replace the infinitesimal $\eta$ in \equ{omega1} with an inverse relaxation time $1/\tau$. In both cases the finite Drude contribution to longitudinal conductivity is
\[ \sigma^{(+)}_{\alpha\beta} = \frac{\tau}{\pi} D_{\alpha\beta} . \] This is clearly reminiscent of the classical Drude theory for free electrons \cite{AM1}, where \[ D = D_{\rm free}\delta_{\alpha\beta}, \quad D_{\rm free} = \frac{\pi e^2 n}{m} . \] One could therefore cast \[ D_{\alpha\beta} = \frac{\pi e^2 n^*_{\alpha\beta}}{m} , \label{n*} \] where $n^*_{\alpha\beta}$
has the meaning of the effective electron density contributing to the macroscopic adiabatic current \cite{Scalapino93}. In the case of a crystalline system of noninteracting electrons only the partly filled bands contribute to $n^*_{\alpha\beta}$ \cite{rap157}; analogously, in the classical theory $n$ is meant to represent the valence electrons only. 
 
\subsection{Born effective charges in metals}

Born effective-charge tensors are a staple in the theory of harmonic lattice dynamics in crystalline insulators, and in the ab-initio theory of ionic conductivity in insulating liquids (molten salts and electrolytes in general). If $s$ labels the $s$-th nucleus in the BvK periodic cell of volume $L^3$ (or in the crystal cell) the total macroscopic  current density flowing through the sample while the nuclei move with velocities ${\bf v}_s$ is \cite{Grasselli19}: \[ j^{\rm(tot)}_\alpha(t) = \frac{1}{L^3} \sum_s e Z^*_{s,\alpha\beta}(t) v_{s\beta}(t) . \label{basic} \]
The expression holds to linear order in the nuclear velocities, and the $Z^*$ tensors depend on time through the instantaneous positions of the nuclear coordinates; in the lattice-dynamical case they are evaluated at the equilibrium crystal structure. Owing to linearity, the dimensionless Born charge tensor at a given time is \[ Z^*_{s,\alpha\beta} = \frac{L^3}{e} \frac{\partial j^{(\rm(tot)}_\alpha}{\partial v_{s\beta}} , \label{def} \] and in insulators the Born tensors obey the basic relationship $\sum_s Z^*_{s,\alpha\beta} =0$, called the acoustic sum rule \cite{PCM}.
  
It was recently discovered by Dreyer, Coh, and Stengel \cite{prep} that, when the definition of \equ{def} is extended to the metallic case, the acoustic sum rule is violated and $\sum_s Z^*_{s,\alpha\beta}$ is proportional to the Drude weight. In order to see how this happens, suppose that all nuclei in the cell are rigidly displaced with the same velocity ${\bf v}$; then \equ{basic} yields \[ j^{\rm(tot)}_\alpha = \frac{e}{L^3} \left(\sum_s Z^*_{s,\alpha\beta} \right) v_{\beta} . \label{basic2} \]  In the insulating case no current flows, i.e. the many-electron system is rigidly translated as well: hence the acoustic sum rule. 

In the metallic case, instead, the electrons are left behind and a macroscopic steady current flows through the sample.
In the reference frame of the nuclei, the macroscopic current is carried by the electrons only, all moving at velocity $- \v$. The same macroscopic current density can then be written as \[ j^{\rm(tot)}_\alpha = e \,n^*_{\alpha\beta} v_\beta, \qquad n^*_{\alpha\beta} = \frac{1}{L^3} \sum_s Z^*_{s,\alpha\beta} , \] where $n^*_{\alpha\beta}$ has the meaning of the effective electron density contributing to the steady current. Comparing to \equ{n*} one immediately gets \[ \frac{1}{L^3} \sum_s Z^*_{s,\alpha\beta} = \frac{m}{\pi e^2} D_{\alpha\beta} .\]  

This result is double checked from the appropriate quantum-mechanical linear responses in Appendix \ref{sec:dcs}; the result requires T-symmetry (and local one-body potential), as indeed recognized in Ref. \cite{prep}.  
  
\section{Quadratic conductivity}

\subsection{Hall conductivity}
The Hall response quadratic in the field obtains by taking the time derivative  of the second term in \equ{current}: 
\bea \partial_t j^{(-)}_\alpha(t) &=& - \frac{e^2}{\hbar L^d}\partial_t \Omega_{\alpha\beta}(\kk) \EE_\beta \\ &=&  - \frac{e^2}{\hbar L^d}\partial_{\kappa_\gamma}\Omega_{\alpha\beta}(0) \EE_\beta \dot\kappa\gamma = \frac{e^3}{\hbar^2 L^d}\partial_{\kappa_\gamma}\Omega_{\alpha\beta}(0) \EE_\beta \EE_\gamma . \nonumber \eea This is clearly constant in time: in absence of any extrinsic relaxation mechanism the many-electron system undergoes a skewed free acceleration: \[ j_\alpha^{(-,2)}(t) = \frac{e^3t}{\hbar^2 L^d}\partial_{\kappa_\gamma}\Omega_{\alpha\beta}(0) \EE_\beta \EE_\gamma.\] The Fourier transform, analogously as for \equ{omega1}, is \[ j^{(-,2)}_\alpha(\omega) = \frac{e^3}{\hbar^2 L^d}\partial_{\kappa_\gamma}\Omega_{\alpha\beta}(0) \frac{i}{\omega + i \eta} \EE_\beta \EE_\gamma ; \label{nhc} \] an alternative derivation is provided in Appendix \ref{sec:freq} directly in the $\omega$ domain.

As for the symmetry properties of the quadratic Hall response, we remind that in presence of T-symmetry $\Omega_{\alpha\beta}(\kk) = - \Omega_{\alpha\beta}(-\kk)$, while in presence of I-symmetry $\Omega_{\alpha\beta}(\kk) = \Omega_{\alpha\beta}(-\kk)$ \cite{Xiao10}: therefore in a T-symmetric system $\Omega_{\alpha\beta}(0)=0$ and---as said above---the linear Hall conductivity vanishes. In the quadratic case the parity is swapped: the gradient of $\Omega_{\alpha\beta}(\kk)$ is even in T-symmetric systems, and odd in I-symmetric systems. Therefore a quadratic Hall current requires breaking of I-symmetry; in the special case of a T-symmetric and I-breaking system, nonzero Hall conductivity appears to second order only. 

In the single-relaxation-time approximation we heuristically replace $\partial_t$  with $\partial_t + 1\tau$ and discard the time-dependent transient current; equivalently we may replace the infinitesimal $\eta$ in \equ{nhc} with $1/\tau$. The quadratic dc current becomes in both cases  \[ j^{(-,2)} = \frac{\tau e^3}{\hbar^2 L^d}\partial_{\kappa_\gamma}\Omega_{\alpha\beta}(0) \EE_\beta \EE_\gamma .\] 

\subsection{Longitudinal conductivity}  

By taking one more time derivative  of the first term in \equ{current} we get: \bea \frac{\partial^2 j^{(+)}_\alpha(t)}{\partial t^2}  &=& - \frac{e}{\hbar L^d} \frac{\partial^3 E_{0\kk}}{\partial t^2 \,\partial \kappa_\beta} \nn &=& - \frac{e}{\hbar L^d} \frac{\partial^3 E_0}{\partial \kappa_\alpha \partial \kappa_\beta \partial\kappa_\gamma} \dot\kappa_\beta \dot\kappa_\gamma \nn &=& - \frac{e^3}{\hbar^3 L^d} \frac{\partial^3 E_0}{\partial \kappa_\alpha \partial \kappa_\beta \partial\kappa_\gamma} \EE_\beta \EE_\gamma. \label{rapix3} \eea In presence of T-symmetry $E_{0\kk} = E_{0,-\kk}$ and therefore \equ{rapix} vanishes: a quadratic contribution to longitudinal conductivity requires T-breaking. 

The second time-derivative of the adiabatic current is time-independent. Therefore  the quadratic response of the many-electron system is a motion where---in absence of dissipation---the acceleration itself increases linearly in time: \[ j^{(+,2)}_\alpha(t)  = - \frac{e^3 t^2}{2 \hbar^3 L^d} \frac{\partial^3 E_0}{\partial \kappa_\alpha \partial \kappa_\beta \partial\kappa_\gamma} \EE_\beta \EE_\gamma . \label{j+} \]
The corresponding expression in the $\omega$ domain is \[ j^{(+,2)}_\alpha(\omega) = - \frac{e^3}{2\hbar^3 L^d} \frac{\partial^3 E_0}{\partial \kappa_\alpha \partial \kappa_\beta \partial\kappa_\gamma} \left(\frac{i}{\omega + i \eta}\right)^2\EE_\beta \EE_\gamma. \label{wata} ,\]  where clearly the singular distribution is the counterpart of $t^2$ dependence of the induced current. To the best of the author's knowledge, this expression first appeared in Ref. \cite{Watanabe20}; an alternative derivation is provided in Appendix \ref{sec:freq}.

In order to heuristically summarize the extrinsic relaxation mechanisms in a single relaxation time, one  may replace the infinitesimal $\eta$ in \equ{wata} with $1/\tau$. The quadratic longitudinal current becomes \[ j^{(+,2)}_\alpha = -\frac{\tau^2 e^3}{2\hbar^3 L^d} \frac{\partial^3 E_0}{\partial \kappa_\alpha \partial \kappa_\beta \partial\kappa_\gamma} ; \label{d3} \] the $\tau^2$ scaling is common to the semiclassical theory \cite{Lahiri21,Tsirkin21,Zhang21}. As said above, in presence of T-symmetry \equ{d3} vanishes and the current quadratic in the field is purely transverse, and is nonzero provided that I-symmetry is broken.

\section{Higher order conductivities}

\subsection{Hall conductivity}

The logic adopted so far extends easily to the nonlinear dc response of any order. In the Hall case the adiabatic current of order $\ell$, for $\ell \ge 2$, obtains form the $(\ell - 1)$-th time derivative  of ${\bf j}^{(-)}(t)$:
\begin{widetext}
\[ \frac{\partial^{\ell-1} j_{\alpha_1}^{(-)}(t)}{\partial t^{\ell-1}} = -\frac{e^2}{\hbar L^d} \left(- \frac{e}{\hbar}\right)^{\ell -1} \frac{\partial^{\ell-1} \Omega_{\alpha_1 \alpha_2}(0)}{\partial \kappa_{\alpha_3}\dots \alpha_{\ell+1}} \EE_{\alpha_2}  \EE_{\alpha_3} \dots \EE_{\alpha_{\ell+1}} .\] Since the $(\ell - 1)$-th derivative is constant in time, the $\ell$-th order adiabatic current evolves in time like $t^{\ell-1}$: \[  j_{\alpha_1}^{(-,\ell)}(t) = - \frac{1}{(\ell-1) !} \frac{e^2}{\hbar L^d} \left(- \frac{et}{\hbar}\right)^{\ell -1} \frac{\partial^{\ell-1} \Omega_{\alpha_1 \alpha_2}(0)}{\partial \kappa_{\alpha_3}\dots \alpha_{\ell+1}}  \EE_{\alpha_2}  \EE_{\alpha_3} \dots \EE_{\alpha_{\ell+1}} , \label{j-} \] and its Fourier transform is:
\[ j^{(-,\ell)}_{\alpha_1}(\omega) = - \frac{1}{(\ell-1)!}\frac{e^2}{\hbar L^d} \left( -\frac{e}{\hbar}\right)^{\ell-1}\frac{\partial^{\ell-1} \Omega_{\alpha_1 \alpha_2}(0)}{\partial \kappa_{\alpha_3}\dots \alpha_{\ell+1}} \left(\frac{i}{\omega + i \eta}\right)^{\ell-1} \EE_{\alpha_2}  \EE_{\alpha_3} \dots \EE_{\alpha_{\ell+1}}\label{nhc2b} ;\]
\end{widetext}
see Appendix \ref{sec:freq} for the derivation of \equ{nhc2b} directly in the $\omega$ domain. The highly singular distribution is once more a fingerprint of the time dependence of the adiabatic current induced by a dc field, which in the present case is $t^{\ell-1}$.
When the infinitesimal $\eta$ is heuristically replaced by a single inverse relaxation time $1/\tau$  all the induced currents become time-independent. The order-$\ell$ Hall currents scale like $\tau^{\ell-1}$, as in the semiclassical theory \cite{Lahiri21,Tsirkin21,Zhang21}.

We remind that in presence of T-symmetry $\Omega_{\alpha\beta}(\kk) = - \Omega_{\alpha\beta}(-\kk)$, while in presence of I-symmetry $\Omega_{\alpha\beta}(\kk) = \Omega_{\alpha\beta}(-\kk)$; therefore the odd-order currents are nonzero only if T-symmetry is broken, while the even-order ones are nonzero only if I-symmetry is broken. If the material is both T-symmetric and I-symmetric no Hall current may flow, to any order in the electric field. Even such features are in agreement with very general symmetry arguments.

\subsection{Longitudinal conductivities}

The adiabatic longitudinal current of order $\ell$ obtains form the $\ell$-th time derivative  of ${\bf j}^{(+)}(t)$ in \equ{current}:
\begin{widetext}
\[ \frac{\partial^{\ell} j^{(+)}_{\alpha_1}(t)}{\partial t^{\ell}} = -\frac{e}{\hbar L^d} \left(- \frac{e}{\hbar}\right)^{\ell} \frac{\partial^{\ell+1} E_0}{\partial  \kappa_{\alpha_{1}} \partial \kappa_{\alpha_2}\dots \partial \kappa_{\alpha_{\ell+1}}} \EE_{\alpha_2} \EE_{\alpha_3} \dots \EE_{\alpha_{\ell+1 }},\] 
Since the $\ell$-th derivative is constant in time, the $\ell$-th order adiabatic current evolves in time like $t^{\ell}$: 
\[  j_{\alpha_1}^{(+,\ell)}(t) = - \frac{1}{\ell\, !} \frac{e}{\hbar L^d} \left(- \frac{et}{\hbar}\right)^{\ell} \frac{\partial^{\ell+1} E_0}{\partial  \kappa_{\alpha_{1}} \partial \kappa_{\alpha_2}\dots \partial \kappa_{\alpha_{\ell+1}}} \EE_{\alpha_2} \EE_{\alpha_3} \dots \EE_{\alpha_{\ell+1 }},\] 
\[  j_\alpha^{(+,\ell)}(\omega) = - \frac{1}{\ell\,!} \frac{e}{\hbar L^d} \left(- \frac{e}{\hbar}\right)^{\ell} \frac{\partial^{\ell+1} E_0}{\partial  \kappa_{\alpha_1} \partial \kappa_{\alpha_2}\dots \partial \kappa_{\alpha_{\ell+1}}} \left(\frac{1}{\omega + i \eta} \right)^{\ell}\EE_{\alpha_2} \EE_{\alpha_3} \dots \EE_{\alpha_{\ell+1} } ; \label{omega3}\] 
\end{widetext}
even here the highly singular distribution is a fingerprint of the $t^{\ell}$ dependence of the adiabatic current induced by a dc field. \equ{omega3} was first obtained by Watanabe and Oshikawa in 2020  \cite{Watanabe20,Watanabe20b}; an alternative derivation is reported in Appendix \ref{sec:freq}; in presence of T-symmetry the odd-order longitudinal conductivities vanish.

When the infinitesimal $\eta$ is heuristically replaced by a single inverse relaxation time $1/\tau$  all the induced currents become time-independent. The order-$\ell$ longitudinal currents scale like $\tau^{\ell}$, as in the semiclassical theory \cite{Lahiri21,Tsirkin21,Zhang21}..

\section{Independent electrons} \label{sec:inde}

\subsection{Band-structure formulation}

I deal next with the special case of band insulators and band metals, i.e. crystalline systems of independent electrons. One needs therefore to express within band-structure theory the two main quantities entering the current as defined in \equ{current}, namely the ground-state energy per unit volume $E_{0\kk}/L^d$ and the many-body curvature per unit volume $\Omega_{\alpha\beta}(\kk)/L^d$.

At the independent-electron level the many-electron wavefunction is a Slater determinant of Bloch orbitals $\ket{\psi_{j\k}} = \ei{\k \cdot \r} \ket{u_{j\k}}$ with band energies $\epsilon_{j\k}$; we normalize the orbitals to one over the crystal cell. The discrete $\k$-vectors become a continuous variable after the $L \rightarrow \infty$ limit is taken \cite{Kittel}.

It is easy to prove (see Appendix \ref{sec:bs})
that \[ \frac{E_{0\kk}}{L^d} = \sum_j \intk f_j(\k) \epsilon_{j,\k\!+\!\kk} , \label{energ} \]
\[ \frac{1}{L^d}\Omega_{\alpha\beta}(\kk) = \sum_j \int_{\rm BZ} \frac{d\k}{(2\pi)^d} f_j(\k) \, \tilde\Omega_{j,\alpha\beta}(\k+\kk) , \label{converg} \] where BZ is the Brillouin zone, and $f_j(\k)$ is the Fermi factor at $T=0$; in \equ{converg} $\tilde\Omega_{j,\alpha\beta}(\k)$ is the Berry curvature of band $j$  \cite{Vanderbilt}:
 \[ \tilde\Omega_{j,\alpha\beta}(\k) = -2\, \mbox{Im } \ev{\partial_{k_\alpha} u_{j\k} | \partial_{k_\beta} u_{j\k}} . \] The formul\ae\ are given per spin channel (or for ``spinless electrons''). 
 
All of the formula\ae\ provided so far in a many-body setting apply as they stand to the independent-electron case, by simply adopting the BZ-integral expressions of \eqs{energ}{converg}; therein---as said above---the large-$L$ limit is implicit. The current induced by a constant field $\EEE$, \equ{current}, becomes in the band-structure case, and  to all orders in $\EEE$: \bea j_\alpha(t) &=& - \frac{e}{\hbar}   \sum_j \intk f_j(\k) \,\partial_{\kappa_\alpha} \epsilon_{j,\k\!+\!\kk}\label{inde}  \\ &-&\frac{e^2}{\hbar}  \sum_j \int_{\rm BZ} \frac{d\k}{(2\pi)^d} f_j(\k) \, \tilde\Omega_{j,\alpha\beta}(\k+\kk) \; \EE_\beta , \nonumber \eea where $\kk(t) = - et\EEE/\hbar$.

The gradient of a function periodical in reciprocal space integrates to zero over the whole BZ. Therefore in insulators all conductivities bar the linear Hall vanish. The alert reader may have noticed that---in the above  many-body formulation---no explicit statement has ruled out nonlinear Hall conductivities in insulators. The reason is that in fact I was unable to reach a proof of this conjecture, the ultimate reason being that discriminating an insulator from a metal is trivial in the band-structure case, much less so in the many-body case \cite{rap_a33}.
As said above, in insulators linear Hall conductivity is quantized for $d=2$; materials realizing the quantum anomalous Hall effect---known as ``Chern insulators'' \cite{Vanderbilt}---have been synthesized since 2013 onwards \cite{Chang13,Chang15}. As a basic tenet of topology, extrinsic effects play no role in such materials, insofar as they remain insulating. The Hall conductivity cannot be quantized for $d=3$, because it has not the dimensions of some fundamental constant.

Switching to metals all contributions from the fully occupied bands vanish, since they integrate to zero over the BZ; only the partially filled bands eventually appear in \equ{inde}. A very hypothetical exception could be a T-breaking metal whose core bands are topological.
Landau's Fermi-liquid theory holds that charge transport in metals involves only quasiparticles with energies within $k_{\rm B}T$ of the Fermi level \cite{Haldane04}, while \eqs{energ}{converg} are instead Fermi-volume integrals. An integration by parts transforms indeed the responses---to any order---into Fermi-surface integrals.
I have tacitly assumed here a simple Fermi surface; in general there may be multiple bands that cross the Fermi level and Fermi surfaces having complex
topology, for example including disconnected sections. Even such cases can be dealt with, but require some extra care (see e.g. Ref. \cite{Wang07}); we disregard such complications here.

The band-structure formulation of the Drude weight is thoroughly discussed in Ref. \cite{rap157}; here I only address the quadratic Hall conductivity, a topic which is drawing a large interest since the seminal 2015 paper by Sodemann and Fu \cite{Sodemann15}, formulated therein within the semiclassical approach.
If we write the quadratic Hall current as \[ j^{(-,2)}_\alpha(\omega) = \frac{i}{\omega +i\eta} \chi_{\alpha\beta\gamma} \EE_\beta \EE_\gamma, \label{chi} \] then \eqs{nhc}{converg} yield \[ \chi_{\alpha\beta\gamma} = \frac{e^3}{\hbar^2} \sum_j \int_{\rm BZ} \frac{d\k}{(2\pi)^d} f_j(\k) \; \partial_{k_\gamma} \tilde\Omega_{j,\alpha\beta}(\k) .  \label{sode} \] 
We have retrieved here the Sodemann-Fu result; see the final part of Sec. \ref{sec:semi} for a further discussion. As usual, the Fermi-volume integral can be transformed in a Fermi-surface integral via an integration by parts. The expression in \equ{chi} also shows a feature recently emphasized in Ref. \cite{Tsirkin21}: the nonlinear conductivity tensors are non unique. In fact addition of an arbitrary term to $\chi_{\alpha\beta\gamma}$, antisymmetric in the $\beta\gamma$ indices, has no effect on the physical current.
 
\subsection{The semiclassical approach} \label{sec:semi}

Many recent papers have considered nonlinear conductivities (longitudinal and transverse) in the framework of semiclassical theory. Common wisdom holds that---whenever dc phenomena are considered---the semiclassical theory provides results which are exact at the band-structure level. To linear order, the longitudinal case is dealt with in Ref. \cite{rap157}, and the Hall case is obvious: the band-structure formula and the semiclassical formula coincide as they stand \cite{Xiao10,Vanderbilt}.
The present formalism leads naturally to a simple proof to an arbitrary order; since the quantum-mechanical response functions imply zero temperature and no dissipation, the semiclassical theory is formulated next under the same conditions.

In order to simplify the algebra we take the simple case of one band and of a Fermi surface which does not touch the BZ boundary. Therefore in \equ{inde} we drop the $j$ index and we replace the BZ integral with the integral over the whole $\k$ space. A change of variables yields \bea j_\alpha(t) &=& - \frac{e}{\hbar}   \int \frac{d\k}{(2\pi)^3} \; f(\k-\kk) \,\partial_{\kappa_\alpha} \epsilon_{\k} \nn &-&\frac{e^2}{\hbar}  \int \frac{d\k}{(2\pi)^d} f(\k-\kk) \, \tilde\Omega_{\alpha\beta}(\k) \; \EE_\beta . \label{semi} \eea This is exact---in the adiabatic approximation---for all times, and we remind once more that $\kk(t) = -e t \EEE/\hbar$; in this alternative form the time dependence of the current originates from the Fermi factor only.

When we expand the Fermi factor in \equ{semi} in powers of the field, the $\ell$-th time-derivative of the $\ell$-th term is constant in time:
 \[\frac{\partial^\ell f(\k)}{\partial t^\ell} = - \left(- \frac{e}{\hbar}\right)^\ell \frac{\partial^{\ell} {f}(\k)}{\partial  k_{\alpha_1} \partial k_{\alpha_2}\dots \partial k_{\alpha_\ell}} \EE_{\alpha_1} \EE_{\alpha_2} \dots \EE_{\alpha_\ell }.\]  
We are ready at this point to make contact with the semiclassical formulation, based on the Boltzmann equation. Therein, the time evolution of the Fermi factor $\tilde f(\k,t)$ is, in the infinite-$\tau$ limit: \[ \partial_t \tilde f(\k,t) = - \dot\k \cdot \nabla_\k \tilde f(\k,t)  . \label{boltz} \] In zero magnetic field the semiclassical equation of motion  for $\k$ is $\dot\k = -e\EEE/\hbar$, ergo $\k(t) =  -et\EEE/\hbar$. Upon deriving $\ell - 1$ times \equ{boltz} one finds the term of order $\ell$ in the field: \bea \frac{\partial^\ell \tilde f(\k,t)}{\partial t^\ell} &=& \frac{e}{\hbar} \frac{\partial^{\ell-1}}{\partial t^{\ell - 1}}\frac{\partial \tilde f(\k,t)}{\partial k_{\alpha_1}} \EE_{\alpha_1} \\= &-& \frac{e}{\hbar} \left(-\frac{e}{\hbar} \right)^{\ell-1} \frac{\partial^{\ell} \tilde{f}(\k,0)}{\partial  k_{\alpha_1} \partial k_{\alpha_2}\dots \partial k_{\alpha_\ell}} \EE_{\alpha_1} \EE_{\alpha_2} \dots \EE_{\alpha_\ell },\nonumber \eea  which is constant in time.
This  shows that the expansion of the zero-temperature semiclassical Fermi factor $\tilde{f}(\k,t)$ in powers of $\EEE$ is identical---for $\tau \rightarrow \infty$---to the expansion of the ab-initio quantum mechanical Fermi factor appearing in \equ{semi}. 

The semiclassical velocity  in zero macroscopic $\B$ field is \[ v_\alpha(\k) = \frac{1}{\hbar} \partial_{\kappa_\alpha} \epsilon_{\kappa_\alpha} + \frac{e}{\hbar} \tilde\Omega_{\alpha\beta}(\k) \EE_\beta , \] and the semiclassical current is \[ j_\alpha(t) = - e  \int \frac{d\k}{(2\pi)^3} \; \tilde f(\k,t) v_\alpha(\k) . \] Therefore, as stated above,  the semiclassical ``approximation'' it is not an approximation after all, insofar as only dc transport---to all orders in $\EEE$---is considered: the semiclassical theory reproduces the exact quantum-mechanical response functions in the framework of band-structure theory, i.e. for a system of noninteracting electrons in a periodic potential. Sch\"rodinger equation obviously implies zero temperature, and the response functions are causal but nondissipative. In a semiclassical approach a relaxation time can be introduced directly within Boltzmann equation, and a finite temperature can be directly accounted for in the form of the Fermi factor $\tilde f(\k,t)$. In the ab-initio approach, instead, these two effects must be accounted for {\it a posteriori} in heuristic ways. 

At input signal of frequency $\omega$ induces---beyond the linear regime---generation of higher harmonics. A macroscopic $\omega$-dependent field induces e.g. to second order a rectified current (time-independent) and a second-harmonic current at frequency $2\omega$: some of the semiclassical literature deals with the different terms separately \cite{Sodemann15,Zhang21}. But since only dc transport is addressed, the $2\omega \rightarrow 0$ limit is eventually taken: only the sum of the two terms is therefore physically relevant. For instance \equ{sode}, in the single-band case, is equivalent to the sum of the two terms in Ref. \cite{Sodemann15}; the present adiabatic derivation avoids partitioning the dc response into different harmonics.

\section{Conclusions}  

I have presented here a comprehensive treatment of nonlinear dc conductivities of any order, Hall and longitudinal, based on Kohn's pathbreaking approach to Drude conductivity. In the Hall case all conductivities are geometrical, in that they are determined by the many-body Berry curvature. At the start, the focus of the theory is on the time dependence of the current adiabatically induced, to a given order, by a static $\EEE$ field in a pristine material. Switching then to the $\omega$ domain, the highly singular Fourier transforms of such currents are regularized in an obvious way by means of a finite relaxation time, analogously to the well known case of linear Drude conductivity \cite{rap157}.

All of the present theory of dc conductivity is formulated by considering the adiabatic response of the many-electron system to a dc field $\EEE$. One might wonder whether this is appropriate to any order in the field: therefore some further comments are in order. First of all, ``adiabatic'' means that a {\it static} perturbation is applied slowly (ideally: infinitely slowly) to the system. The response can then be obtained from the (time evolution of) the instantaneous adiabatic ground eigenstate: no perturbation theory is needed.
In the present case the physical perturbation is clearly static: a dc electric field, and therefore it does not violate the adiabatic requirement to all orders. 

The problem is that quantum mechanics---at variance with semiclassical theories---cannot deal with fields directly: it deals with {\it potentials}: vector potential and scalar potential. The latter (in our case $-\EEE \cdot \r$) is static, but it is incompatible with the BvK boundary conditions of condensed matter physics \cite{rap100}. In the case of dc conductivity the requirement of dealing with an unbounded system within BvK boundary conditions is even more stringent: no dc current may flow across a bounded sample isolated in vacuo. Incidentally it is worth mentioning that a  vestige of the Drude weight can nonetheless be retrieved even in a  (bounded) metallic crystallite \cite{rap160}. 

For the above reasons it is mandatory---when dealing with dc conductivity---to adopt the time-dependent vector potential gauge, as done here throughout: in the notations of the present work, the field enters Schr\"odinger equation via the vector potential  $-ct\EEE = \hbar c \kk(t)/e$. The adiabatic response to this perturbation is evaluated throughout to all orders.

The final part of this work addresses a crystalline system of noninteracting electrons and reformulates the whole theory in that framework, where the nonlinear conductivities assume the form of Fermi-volume integrals, or equivalently of Fermi-surface integrals. Finally, the common wisdom that---insofar as dc conductivity is addressed---the semiclassical treatment is exact at the band-structure level is confirmed to all orders in $\EEE$.

\section*{Acknowledgments} 

I thank Ivo Souza for illuminating discussions over many years, and for bringing some relevant papers to my attention. Work supported by the Office of Naval Research (USA) Grant No. N00014-20-1-2847.

\appendix

\section{Kubo formul\ae\ for linear conductivity}

\subsection{Many-body formul\ae}\label{sec:kubo}

We define the matrix elements of the many-body velocity operator at $\kk=0$ :
 \bea {\cal R}_{n,\alpha\beta} &=& \mbox{Re }\langle \Psi_0 | \hat v_\alpha | \Psi_n \rangle \langle
\Psi_n | \hat v_\beta | \Psi_0 \rangle  , \\  {\cal I}_{n,\alpha\beta} &=& \mbox{Im }\langle \Psi_0 | \hat v_\alpha | \Psi_n \rangle \langle \Psi_n | \hat v_\beta | \Psi_0 \rangle ,
\label{vmat} \eea where ${\cal R}_{n,\alpha\beta}$ is symmetric and ${\cal I}_{n,\alpha\beta} $ antisymmetric; we further set $\omega_{0n} = (E_n - E_0)/\hbar$.
The longitudinal (symmetric) conductivity is: \[  \sigma_{\alpha\beta}^{(+)}(\omega) = D_{\alpha\beta} \left[ \delta(\omega) + \frac{i}{\pi \omega} \right] +\sigma_{\alpha\beta}^{(\rm regular)}(\omega) , \label{cond} \] 
\index{conductivity}
\[ D_{\alpha\beta} = \frac{\pi e^2}{ L^d} \left( \frac{N}{m} \delta_{\alpha\beta} - \frac{2}{\hbar} {\sum_{n\neq 0}}  \frac{{\cal R}_{n,\alpha\beta}  }{\omega_{0n}} \right) \label{drude} , \]
\bea \mbox{Re } \sigma_{\alpha\beta}^{(\rm regular)}(\omega) &=& \frac{\pi e^2}{\hbar L^d}  {\sum_{n\neq 0}} \frac{ {\cal R}_{n,\alpha\beta}}{\omega_{0n}} \delta(\omega - \omega_{0n}) , \; \omega>0 \label{s1} \\ \mbox{Im } \sigma_{\alpha\beta}^{(\rm regular)}(\omega) &=& \frac{2 e^2}{\hbar L^d}  {\sum_{n\neq 0}} \frac{ {\cal R}_{n,\alpha\beta}}{\omega_{0n}} \frac{\omega}{\omega_{0n}^2 - \omega^2} \label{s2} ; \eea  the Drude weight $D_{\alpha\beta}$ vanishes in insulators. 

The real part of longitudinal conductivity obeys the $f$-sum rule  \bea \int_0^\infty d \omega \; \mbox{Re } \sigma_{\alpha\beta} (\omega) &=& \frac{D_{\alpha\beta}}{2} + \int_0^\infty d \omega \; \mbox{Re } \sigma_{\alpha\beta}^{(\rm regular)} (\omega) \nn &=& \frac{D_{\rm free}}{2} \delta_{\alpha\beta} , \label{fsum}  \eea  with $D_{\rm free} = \pi e^2 n/m$. An important subtlety must be stressed \cite{Allen06}: if the $\omega$-integration includes the ultraviolet and x-ray regions of the spectrum, then the density $n$ includes the core electrons. When the focus is on dc transport and optical properties the core contributions to the $f$-sum rule must be discounted: this happens automatically in a pseudopotential framework.

Using the relationship \[ \ket{\partial_{\kappa_\alpha} \Psi_0} = - \sum_{n \neq 0} \ket{\Psi_n} \frac{\me{\Psi_n}{\hat v_\alpha }{\Psi_0}}{\omega_{0n}} , \label{der} \]
the Drude weight  can be recast as a geometrical property of the electronic ground state: \[  D_{\alpha\beta} = D_{\rm free} \delta_{\alpha\beta} - \frac{2 \pi e^2}{\hbar^2 L^d} \mbox{Re } \me{\partial_{\kappa_\alpha} \Psi_0}{\,( \hat{H} - E_0 )\,}{\partial_{\kappa_\beta} \Psi_0} .  \label{geom} \] If we then start from the identity $ \me{\Psi_{0\kk}}{\,( \hat{H}_{\kk} - E_{0\kk} )\,}{\Psi_{0\kk}} \equiv 0 \label{iden} $, we take two derivatives, and we set  $\kk = 0$, we arrive at Kohn's famous expression for the Drude weight:  \[ D_{\alpha\beta} = \frac{\pi e^2}{\hbar^2 L^d} \frac{\partial^2 E_0}{\partial \kappa_\alpha \partial \kappa_\beta} , \] also proved in the main text and in Appendix \ref{sec:freq} in two alternative ways.

Transverse conductivity is nonzero only when T-symmetry is absent. The Kubo formul\ae\  for the transverse (antiymmetric) conductivity are:
\bea \mbox{Re } \sigma_{\alpha\beta}^{(-)}(\omega) &=& \frac{2e^2}{\hbar  L^d} {\sum_{n\neq 0}} \frac{{\cal I}_{n,\alpha\beta}}{\omega_{0n}^2 - \omega^2} \label{s3}
\\ \mbox{Im } \sigma_{\alpha\beta}^{(-)}(\omega) &=& \frac{\pi e^2}{\hbar L^d} {\sum_{n
\neq 0}} \frac{{\cal I}_{n,\alpha\beta}}{\omega_{0n}} \delta(\omega - \omega_{0n}) , \; \omega> 0 . \label{s4} \eea 
Using again \equ{der} the dc transverse conductivity is easily recast in terms of the many-body Berry curvature at $\kk=0$: \[ \mbox{Re } \sigma_{\alpha\beta}^{(-)}(0) = - \frac{e^2}{\hbar  L^d}  \Omega_{\alpha\beta}(0) ; \label{main} \] the expression holds for metals and insulators, in either 2$d$ or 3$d$. Notice that $\mbox{Re } \sigma_{\alpha\beta}^{(-)}(0) = \sigma_{\alpha\beta}^{(-)}(0)$, since the imaginary part is odd in $\omega$. \equ{main} is derived in the main text in an alternative way, by means of the anomalous velocity in its many-body formulation. 

\subsection{Band-structure formula\ae} \label{sec:bs} 

We start reminding that the anomalous Hall conductivity of a pristine crystal is \cite{Vanderbilt} \[  \sigma_{\alpha\beta}^{(-)}(0) = - \frac{e^2}{\hbar} \sum_j \int_{\rm BZ} \frac{d\k}{(2\pi)^d} f_j(\k) \, \tilde\Omega_{j,\alpha\beta}(\k) ; \]  comparison to \equ{main} yields: \[ \frac{1}{L^d} \Omega_{\alpha\beta}(0) = \sum_j \int_{\rm BZ} \frac{d\k}{(2\pi)^d} f_j(\k) \, \tilde\Omega_{j,\alpha\beta}(\k) . \] The $L \rightarrow \infty$ is implicitly understood in the l.h.s.; it is instead explicit in the r.h.s., given that the Bloch vector therein is a continuous variable.

When $\kk \neq 0$ is set in Kohn's Hamiltonian $\hat{H}_{\kk}$, the corresponding Kohn-Sham periodic orbitals $\ket{u_{j\k}}$ are eigenstates of the single-particle Hamiltonian \[ \emi{\k \cdot \r}H_{\kk}\ei{\k \cdot \r} = 
\frac{1}{2m} 
\left[ \p + \frac{e}{c}{\bf A}(\r) + \hbar \k + \hbar \kk \right]^2 \!\!+ V_{\rm KS}(\r) , \] where $V_{\rm KS}$ is the Kohn-Sham potential, hence \equ{energ} in the main text is obvious, while for the Berry curvature we get \[ \frac{1}{L^d}  \Omega_{\alpha\beta}(\kk) = \sum_j \int_{\rm BZ} \frac{d\k}{(2\pi)^d} f_j(\k) \, \tilde\Omega_{j,\alpha\beta}(\k+ \kk) . \] \[ \frac{1}{L^d}\partial_{\kappa_\alpha} \Omega_{\alpha\beta}(0) = \sum_j \int_{\rm BZ} \frac{d\k}{(2\pi)^d} f_j(\k) \; \partial_{k_\alpha} \tilde\Omega_{j,\alpha\beta}(\k) . \]   

\section{The Dreyer-Coh-Stengel sum rule} \label{sec:dcs}

When the nuclei are displaced by ${\bf u}_s$ from a reference configuration, the potential in \equ{kohn}---one-body term thereof---depends on such displacements: $\hat V \rightarrow \hat V(\{{\bf u}_s\})$; the ground-state eigenstate $\ket{\Psi_0}$ depends on the ${\bf u}_s$ as well. The electronic current density induced when the Hamiltonian is adiabatically varied in time---using once more the Niu-Thouless theorem \cite{Niu84,Xiao10}---and setting $\kk=0$ is \bea j_\alpha &=& \frac{ie}{L^3}( \ev{\partial_{\kappa_\alpha}{\Psi}_{0} |  \dot{\Psi}_{0} } - \ev{\dot{\Psi}_{0} | \partial_{\kappa_\alpha} \Psi_{0} } )  \nn&=& - \frac{2ie}{L^3} \mbox{Im } \ev{\partial_{\kappa_\alpha}{\Psi}_{0} |  \dot{\Psi}_{0} }, \eea where we recognize a many-body Berry curvature in the $(\kappa_\alpha,t)$ domain. If only the $s$-th nucleus is displaced with velocity ${\bf v}_s = \dot{\bf u}_s$, the electronic current and the corresponding $s$-th Born tensor are   \[ j_{s\alpha} = -\frac{2ie}{L^3}\mbox{Im } \ev{\partial_{\kappa_\alpha}{\Psi}_{0} |  \partial_{u_{s\beta}} {\Psi}_{0} } \, v_{s\beta}  , \] 
\[ Z^*_{s,\alpha\beta} = \frac{1}{e}\frac{\partial j_{\alpha}^{\rm(tot)}}{\partial v_{s,\beta}} = Z_s\delta_{\alpha\beta}- \frac{2i}{L^3} \mbox{Im } \ev{\partial_{\kappa_\alpha}{\Psi}_{0} |  \partial_{u_{s\beta}} {\Psi}_{0} }  , \] where $e Z_s$ is the bare nuclear charge.

If all the nuclei are rigidly translated at velocity ${\bf v}$ the adiabatic electronic current  is \[ j_{\alpha} =   -\frac{2ie}{L^3}\mbox{Im } \ev{\partial_{\kappa_\alpha}{\Psi}_{0} |  \partial_{u_{\beta}} {\Psi}_{0} } \, v_{\beta} ; \label{dcs1} \] the main object of the present Appendix becomes then \bea \frac{1}{L^3} \sum_s Z^*_{s,\alpha\beta} &=& \frac{1}{L^3}\sum_s Z_s \delta_{\alpha\beta} + \frac{1}{e} \frac{\partial j_\alpha}{\partial v_\beta} \label{rap}\\ &=& \frac{m}{\pi e^2} D_{\rm free} \delta_{\alpha\beta} -\frac{2i}{L^3} \mbox{Im } \ev{\partial_{\kappa_\alpha}{\Psi}_{0} |  \partial_{u_{\beta}} {\Psi}_{0} } , \nonumber \eea where $D_{\rm free} = \pi e^2 n/m$. and $n =\sum_s Z_s/L^3$.

Then, for a T-invariant system we may transform \[ \ket{\partial_{\bf u} \Psi_0} = - \sum_{i=1}^N \ket{\partial_{\r_i} \Psi_0} = \frac{i}{\hbar} \sum_{i=1}^N {\bf p}_i \ket{\Psi_0} = \frac{im}{\hbar^2} (\partial_{\kk} \hat H_{\kk}) \ket{\Psi_0}.\] We further exploit 
\bea (\partial_{\kk} \hat{H}_{\kk}) \ket{\Psi_{0\kk}} &=& \partial_{\kk} ( \, \hat{H}_{\kk} \ket{\Psi_{0\kk}} \,) -  \hat{H}_{\kk} \ket{\partial_{\kk} \Psi_{0\kk}} \\ &=& ( \partial_{\kk}\E_{0\kk}) \, \ket{\Psi_{0\kk}}  + (E_{0\kk} -  \hat{H}_{\kk}) \, \ket{\partial_{\kk} \Psi_{0\kk}} . \nonumber \eea 
The second term in the second line does not contribute to \equ{dcs1}; the electronic current induced by a rigid translation of all nuclei is then  \[ \frac{1}{e}\frac{\partial j_{\alpha}}{\partial v_\beta}  =  -\frac{2m}{h^2 L^3}\mbox{Re } \ev{\partial_{\kappa_\alpha}{\Psi}_{0} | \, ( \hat H - E_{0}  ) \, |\partial_{u_{\beta}} {\Psi}_{0} }   \label{dcs2} ;\] inserting this into \equ{rap} and comparing to the geometrical expression for the Drude weight, \equ{geom}, the Dreyer-Coh-Stengel sum rule \cite{prep} is finally retrieved.

\section{Adiabatic response in the  frequency domain} \label{sec:freq}

From $\dot\kk(t) = - e\EEE/\hbar$ one gets $\kk(t) = - -et\EEE/\hbar + \mbox{const}$, and we observe that the Fourier transform of a constant is proportional to $\delta(\omega)$. Switching to the Fourier transforms $\EEE(\omega) = i\omega \hbar \kk(\omega)/e$, whose inversion is \[ \kk(\omega) = \left(\frac{e}{\hbar}\right)\left[-\frac{i}{\omega} + \mbox{const} \times \delta(\omega) \right]\EEE(\omega) ; \] the integration constant, as usual, is determined by imposing causality: \bea \frac{\partial \kappa_\alpha(\omega)}{\partial \EE_\beta(\omega)}  &=& \left(-\frac{e}{\hbar}\right) \left(\frac{i}{\omega + i\eta}\right) \delta_{\alpha\beta}\nn &=& \left(-\frac{e}{\hbar}\right) \left[\frac{i}{\omega} + \pi \delta(\omega) \right] \delta_{\alpha\beta} , \eea where $\eta \rightarrow 0^+$ is understood (as throughout this paper).

The linear conductivity is \bea \sigma_{\alpha\beta}(\omega) &=& \frac{\partial j_\alpha(\omega)}{\partial \EE_\beta(\omega)} \nn &=& \left(-\frac{e}{\hbar}\right) \frac{\partial j_\alpha(\omega)}{\partial \kappa_\beta(\omega)}  \frac{i}{\omega + i\eta} .\eea At finite frequency, $\partial j_\alpha(\omega)/\partial \kappa_\beta(\omega)$ obtains  from time-dependent linear response theory (i.e. Kubo formul\ae). Here we limit ourselvess to an adiabatic perturbation, hence the $\kk$-derivative is taken with respect to a static $\kk$. We thus get from \equ{current}  the dc contribution to longitudinal conductivity as \cite{Scalapino93}: \bea \sigma^{(+)}_{\alpha\beta}(\omega) &=& \left(-\frac{e}{\hbar}\right) \frac{\partial j^{(+)}_\alpha}{\partial \kappa_\beta}  \frac{i}{\omega + i\eta} \nn &=& \frac{e^2}{\hbar^2 L^d} \frac{\partial^2 E_0}{\partial \kappa_\alpha\partial \kappa_\beta} \frac{i}{\omega + i\eta} ,\eea which in fact is Kohn's expression, \eqs{rapix2}{omega1}.

To higher orders in $\EEE$ the adiabatic response of the many-electron system obtains from the chain rule:
\begin{widetext} 
\[  j_{\alpha_1}^{(+,\ell)}(\omega) = - \frac{1}{\ell\,!} \frac{e}{\hbar L^d} \left(- \frac{e}{\hbar}\right)^{\ell} \frac{\partial^{\ell+1} E_0}{\partial  \kappa_{\alpha_{1}} \partial \kappa_{\alpha_2}\dots \partial \kappa_{\alpha_{\ell+1}}} \left(\frac{1}{\omega + i \eta} \right)^{\ell}\EE_{\alpha_2} \EE_{\alpha_3} \dots \EE_{\alpha_{\ell+1} } ; \label{omega3b}\] with the heuristic substitution $\eta = 1/\tau$ the steady current scales like $\tau^\ell$, in agreement with the semiclassical theories.

The Hall current ${\bf j}^{(-,\ell)}$ has been derived in the main text for $\ell=1$ and $\ell=2$; it is expedient to rewrite ${\bf j}^{(-,2)}$, \equ{nhc}, as  \[ j^{(-,2)}_{\alpha_1}(\omega) = - \frac{e^3}{\hbar^2 L^d}\partial_{\kappa_{\alpha_3}}\Omega_{\alpha_1 \alpha_2}(0) \frac{i}{\omega + i \eta} \EE_{\alpha_2} \EE_{\alpha_3} .  .\] Using as above the chain rule, the generalized formula for $\ell \ge 2$ is \[ j^{(-,\ell)}_{\alpha_1}(\omega) = - \frac{1}{(\ell-1)!}\frac{e^2}{\hbar L^d} \left( -\frac{e}{\hbar}\right)^{\ell-1}\frac{\partial^{\ell-1} \Omega_{\alpha_1 \alpha_2}(0)}{\partial \kappa_{\alpha_3}\dots \alpha_{\ell+1}} \left(\frac{i}{\omega + i \eta}\right)^{\ell-1} \EE_{\alpha_2}  \EE_{\alpha_3} \dots \EE_{\alpha_{\ell+1}} .\]
The above currents, where singular distributions appear, are the causal Fourier transforms of the corresponding time-dependent currents, as displayed in the main text.
\end{widetext}


\end{document}